\documentclass[useAMS,usenatbib]{mn2e}

\usepackage{amssymb}
\usepackage{graphicx}
\usepackage{color}
\usepackage{BibDef}
\def\lsim{\mathrel{\rlap{\lower 3pt \hbox{$\sim$}} \raise 2.0pt \hbox{$<$}}}
\def\gsim{\mathrel{\rlap{\lower 3pt \hbox{$\sim$}} \raise 2.0pt \hbox{$>$}}}
\def\msun{\rm {M_\odot}}
\def\kms{\rm km\,s^{-1}}

\title[Constraining the high redshift formation of black hole seeds] 
{Constraining the high redshift formation of black hole seeds in nuclear star clusters with gas inflows}
\author[A. Lupi et al.]{A. Lupi $^{1}$\thanks{E-mail:
alessandro.lupi@uninsubria.it}, M. Colpi$^{2}$, B. Devecchi, 
G. Galanti$^1$, and M. Volonteri$^3$\\
$^1$DiSAT, Universit\`a degli Studi dell'Insubria, Via Valleggio 11, I-22100 Como, Italy\\
$^2$INFN, Sezione di Milano Bicocca, Piazza della Scienza 3,  I-20126 Milano, Italy\\\
$^3$Institut d'Astrophysique de Paris, 98bis Bd. Arago, 75014, Paris, France}
\begin{document}

\date{Accepted 2014 June 4}

\pagerange{\pageref{firstpage}--\pageref{lastpage}} \pubyear{2014}

\maketitle

\label{firstpage}

\begin{abstract}
In this paper we explore a possible route of black hole seed formation that appeal to a model by Davies, Miller \& Bellovary
who considered the case of the dynamical collapse of a dense cluster of stellar black holes subjected to an inflow of gas. 
Here, we explore this case in a broad cosmological context. 
The working hypotheses are that (i) nuclear star clusters form at high redshifts in pre-galactic discs hosted in dark matter halos, providing a suitable environment for the formation of stellar black holes in their cores,  (ii)  major central inflows of gas occur onto these clusters due to
instabilities seeded in the growing discs and/or to mergers with other gas-rich halos, and that (iii) following the inflow,  
stellar black holes in the core avoid ejection due to the steepening to the potential well, leading to core collapse  and the formation of a massive seed of $\lsim 1000\,\msun$.  We simulate a cosmological box tracing the build up of the dark matter halos and there embedded baryons, and
explore cluster evolution with a semi-analytical model.
We show that this route is feasible, peaks at redshifts $z\lsim10$ and occurs in concomitance 
with the formation of seeds from other channels. The channel is competitive relative to others,  
and is independent of the metal content of the parent cluster.  This mechanism of gas driven core collapse requires inflows
with masses at least ten times larger than the mass of the parent star cluster, occurring on timescales shorter than the evaporation/ejection time of the stellar black holes from the core. In this respect,
the results provide upper limit to the frequency of this process.

\end{abstract}
\begin{keywords}
stars: black holes - black hole physics - galaxies: formation - galaxies: evolution.
\end{keywords}

\section{Introduction}

Observations of high redshift quasars show  that black holes as massive as $\gsim 10^9\rm M_{\odot}$ were already in place at redshift $z\lsim 7$ 
\citep{mort11} when the universe was only eight hundred million years old.
It is thus challenging to understand how these black holes formed 
when the Universe was less than $1$ Gyr old and galaxies were in the process of forming.

Current models suggest that 
supermassive black holes  may have formed from {\it seeds} of yet unknown mass which later grew via 
sustained accretion at critical or super-critical rates, and via mergers with other 
black holes during the hierarchical assembly of their host halos \citep{volonterireview10}. 
The origin and nature of this seed population(s) remain uncertain as uncertain are the
physical mechanisms at play.  This has led to speculate that black hole seeds had masses in a wide range,
between $100\,\msun$ - $10^6\,\msun.$ To ensure early formation, seed black holes must have formed in the most massive and rare halos of a given epoch which virialise and grow in the knots of the cosmic web \citep{dimatteo08}.

Seed black holes may have formed as early as redshift $z\sim 20$,  from the core  collapse 
of the first  very massive stars ($\gsim 260\,\msun$) formed out of pristine gas clouds fragmenting in virialized 
pre-galactic halos of $10^{5-6}\,\msun$, i.e. the Pop III stars \citep{haiman96,
tegmark97,heger03, madau01}. The lack of metals suggest a top-heavy initial mass function and thus  a first generation of
seed black holes of stellar origin with masses $\lsim 10^3\,\msun$ \citep{abel02,omukai01,bromm02,bromm03,yoshida08}.
But, recent studies seem to revise the initial estimates of the stellar masses to possibly much lower values of just a few tens of solar masses \citep{clark11,greif11,wise12}.

Heavy seeds can form at later times ($z\lesssim 12$)  from the central collapse of gas in unstable proto-galactic discs of $10^{5-6}\,\msun$ present in heavier ($10^{7-8}\msun$) dark matter halos \citep{koushiappas04,begel06,lodato06}. If the angular momentum barrier and 
the process of fragmentation are suppressed in some of these pre-galactic structures, due to e.g. the presence of a background of UV radiation 
\citep{agarwal12,latif13},  the infall of gas can continue unimpeded leading to the formation of seed black holes by direct collapse.
Alternatively, a non-thermally relaxed giant star can form in these halos which is able to grow a black hole (of $\lsim 10\,\msun$) in its centre \citep{begel10,choi13}. 
In this last case, subsequent super-Eddington growth of this embryo black hole continues, inside the optically thick hydrostatic
envelope, called ``quasi-star'' \citep{begel08}. 
Seed black holes as heavy as $10^{4-5}\,\msun$ can form, depending on the extent of the radiative feedback and metal content  \citep{montero12,dotan11}. 
A further path of black hole seed formation, not constrained by limits imposed by the
metallicity content, calls for gas-rich galaxy mergers which trigger huge inflows of gas on sub-parsec scales  \citep{mayer10}. But avoiding fragmentation of the growing gas cloud due to cooling is again matter of debate \citep{ferrara13}.

An alternate route for seeds formation (up to $10^2\,\msun-10^3\,\msun$) has been explored considering  
galactic discs enriched above a critical metallicity ($Z_{\rm crit}\sim10^{-5}Z_\odot$) and in which cloud fragmentation lead to ordinary Pop II star formation (\citeauthor{dev09}, 2009; D09 hereon). 
This path considers the formation, in some halo, of a young ultra-dense nuclear star cluster (NSC, hereafter) dominated by stellar collisions and
mass segregation in its centre.  A very massive star forms via star-star runway collisions in the cluster core before single stars have time to explode in supernovae. The ensuing collapse  of the runaway star eventually leads to the formation of a black hole above $\gsim 10^{2-3} \,\msun$ \citep{portegies02,gurkan04,dev09}. \footnote{This pathway holds in enriched, yet metal-poor star clusters (those with metallicity below $\sim 10^{-3}\,Z_\odot$), as massive metal enriched stars lose mass via intense wind during nuclear evolution leaving a lower mass black hole or a
neutron star \citep{heger03}. 
 Mergers of metal rich stars can also be very disruptive \citep{glebbeek09}.}
In a cosmological context, this route has been explored by D09
who considered seed formation in pre-galactic discs. It was shown that NSCs of $10^{5-6}\,\msun$ develop as early as $z\sim 15-10$ and 
that the first seed black holes, from runaway stars, start dominating below this redshifts, over the Pop III channel active at earlier epochs (D09, \citeauthor{dev10}, 2010;
\citeauthor{dev12}, 2012; D10, D12 hereon, respectively). 

\subsection{The Gas-Induced Runaway Merger model: GIRM} 
None of the above mentioned mechanisms consider the formation of massive black hole seeds from 
the runaway merger of {\it stellar} black holes in ultra
dense star clusters.  This channel was studied by  Quinlan \& Shapiro in the late eighties and has been revisited recently  by 
\citet{davies11} who introduced a variant to this model.

In \citet{quinlan87}, a very massive Newtonian cluster of compact stars (neutron stars and stellar black holes) 
evolves dynamically into a state of catastrophic core collapse. 
Under the rather extreme conditions considered by Quinlan \& Shapiro ( i.e. of a galactic nucleus of $10^{7-8}$ stars with high initial 
velocity dispersion $\gsim 1000\,\kms$)  binaries of compact stars form mainly via dissipative two-body encounters under
the control of gravitational wave emission, ending in their coalescence.
Following unimpeded self-similar core collapse, the central density, gravitational potential and velocity dispersion rise up to
a critical limit
corresponding to the onset of 
a relativistic dynamical instability (when the gravitational redshift $z_{\rm grav}$ rises above  $\sim 0.5$). 
A central 
 $\gsim100\,\msun$ black hole forms embracing the mass of the collapsing core.

In a subsequent paper, \citet{quinlan89} recognised that the inclusion of a realistic mass spectrum for the compact stars combined with 
an accurate modelling of the stellar dynamics prevents the core from reaching 
the high velocity dispersions requested for the onset of the relativistic dynamical collapse. 
Fast mass segregation by dynamical friction combined with the tendency of
stars to reach equipartition of kinetic energy (i.e. lower dispersion velocities for the heavier stars), led the core to evolve into a state of declining velocity dispersion, preventing the deepening of the gravitational potential well with time up to $z_{\rm grav}\sim 0.5$ (see e.g. figure 3 of Quinlan \& Shapiro 1989). 
The ultimate fate of the stellar  
black holes left in the centre was then found very difficult to calculate and predict, within the limits imposed by the Fokker Planck approximation used to study the system. \footnote{Note that in a follow up paper, \citet{quinlan90} showed that in galactic nuclei there is the natural tendency of 
triggering at their centre runway star-star collisions during the early stage of their evolution. The very massive star 
is then expected to collapse into a massive black hole. This channel was later re-proposed by \citet{portegies02} in the context of young 
and dense star clusters of lower mass after the recognition that a wide mass spectrum leads to rapid mass segregation and multiple 
stellar runaway collisions ending with the formation of a massive object of $100\,\msun$.}

\citet{quinlan87} first recognised that only in massive galactic nuclei of $10^{7-8}$ stars with velocity dispersions in excess of $1000\,\kms$, 
binary stars, known to act as a kinetic energy source, do not heat dynamically the cluster to halt and reverse the gravo-thermal 
catastrophe of the core.
In lower mass systems, such as globular clusters comprising $10^{5-6}$ stars, recent numerical studies indicate that most of the
stellar black holes born from ordinary stellar evolution, either single or in binaries, are ejected from the cluster after a few Gyrs from formation. 
This follows in response to close single-binary interactions which harden the black hole binaries imparting large recoil
velocities, and/or to binary coalescences via gravitational wave emission for which the induced recoil velocity can largely
exceed the escape speed from the cluster \citep{downing11,oleary06}.  Either a single or a binary stellar black hole  or no black hole is left at the centre, after ejection of the bulk of the population.\footnote{There are theoretical claims that stellar mass black holes are present
in some globular cluster \citep{moody09} and a recent observation seems to support this view \citep{strader12}.}

\citeauthor{davies11} (2011; DMB11 hereafter) reconsider this channel of runaway merging of stellar black holes
in dense star clusters, in a broader context.  They envisage 
the case of a galaxy undergoing a merger with another galaxy or the case of a large-scale
gravitational instability in a galaxy which drives (in both cases)  
a major central inflow of gas onto a pre-existing NSC, i.e. an inflow that can even exceed the mass of the star cluster itself. 
In the case the in-falling gas dominates the cluster gravitational potential, the dynamical behaviour of all stars in the cluster
changes in response to the inevitable gas-induced increase of the dispersion velocity.
Hard binaries present in the cluster (i.e. binaries which carry a binding energy per unit mass larger than the kinetic energy 
per unit mass of single stars) are turned nominally into soft binaries which soften when interacting with single stars. 
This process has the effect of reducing  the dynamical heating of the cluster, now more
susceptible to the gravo-thermal collapse. 
The cluster, dominated by the gravity of the underlying gas, may than enter  
a period of core collapse: stellar mass black holes in the hardest binaries 
start merging via gravitational radiation reaction before heating the cluster and/or being dynamically ejected.
The enhanced escape speed due to the inflow reduces the effect of black hole ejection.
Eventually
the black holes in the core merge in a runaway fashion. 
Central velocities as large as $\gsim 1000 \,\kms$ needs to be attained following a major gas inflow (see DMB11). 

The details of this model, and in particular the effect of an inflow of gas on the fate of the stellar black holes,  
have not been exploited yet and future dedicated numerical experiments are necessary to asses the feasibility of this route.
Hereon, we will assume that this channel operates
in NSC subjected to a major inflow, postponing to an incoming paper the analysis of the interaction of the gas with the stars and black holes 
in the cluster (Galanti et al. in preparation). As large central  inflows of gas are expected to occur during the build up of pre-galactic structures, 
this mechanism for black hole seed formation may be relevant during the early phases of galaxy formation.

Aim of this work is at exploring whether major inflows of gas occur at the centre of pre-galactic disc hosting 
a NSC, to asses the cosmological relevance of this process under the hypothesis that inflows of gas drive
 the runaway merger of the stellar black holes in these clusters. 
To this purpose, we study
how a population of NSCs form and evolve 
inside dark matter halos assembling out to redshifts as large as $\sim 20$, in concordance  with the $\Lambda$CDM paradigm for hierarchical 
structure formation. 
To achieve this goal, we evolve  ``Pinocchio'' \citep{monaco02}, a code  which follows the cosmological evolution of dark matter halos with cosmic time, and a second code by D09, D10, D12 which creates and follows the evolution of the baryonic components in the halos, using 
physically motivated prescriptions for disc and star formation in pre-galactic halos.
D09, D10, D12 studied a scenario of black hole seed formation that jointly account for 
the early formation of a population of Pop III  stars with their relic black holes,  and at later times of young NSCs able to grow in their core a 
runaway star resulting from repeated stellar collisions which is fated to become a black hole of large mass. 
 D10 and D12 select clusters having relaxation times shorter than the evolution time of the massive stars ($\lsim10^6$ yr) for this to happen. 

In this paper, we modify the code in order to include the Gas Induced  black hole Runaway Merger model 
(hereon GIRM model) which is replacing  the stellar-runaway NSC channel.
The aim is at determining when black hole seeds form via the GIRM,  
tracing self-consistently the cosmological gas inflows in galactic halos. 
This implies the selection of NSCs which
undergo ordinary stellar evolution and  are subject to  major gas inflows during the cosmic assembly of structures. 
We note that the stellar-runaway NSC channel is complementary to the GIRM model as the latter
evolves stars under ordinary conditions. Thus a comparative analysis will enable us to asses the importance of GIRM, relative to Pop III
and the stellar-runaway channel, in a cosmological 
context.
In Section \ref{sec:DMhalo}, we describe shortly the formation and assembly of dark matter halos and their embedded NSCs, while 
in Section \ref{sec:gasdriven} we outline the input physics of the GIRM model, deferring the reader to D10 and D12 for details. 
Section \ref{sec:results} illustrates the results and Section \ref{sec:conclusions} contains our conclusions.

\section{Dark matter halos, galactic discs and the formation of nuclear star clusters}
\label{sec:DMhalo}
``Pinocchio''  is a code \citep{monaco02} which evolves an initial density perturbation field, on a 3D grid, 
using the Lagrangian Perturbation Theory in order to generate catalogues of virialized  halos  at different cosmic times,
keeping track of the halo's mass, position, velocity, spin parameter, and merger history. 
In this paper we evolve a cosmological volume of 10 Mpc  in co-moving length and adopt  the $\Lambda$CDM cosmology with 
$\Omega_{\rm baryon} =0.041$, $\Omega_{\rm matter} =0.258$, $\Omega_{\Lambda}=0.742$, $h=0.742$, and $n_s= 0.963$
\citep{dunkley09}. 

In this simulation the mass resolution is of $M_{\rm h,min}=3\times 10^5\,\msun$  for the dark matter halos, and is computed as the total mass in the box divided by the number of elements assigned at the
onset of the simulation \citep{monaco02}. 
Dark matter elements  lighter than $M_{\rm h,min}$ do not form in the box.
The corresponding resolution limit for baryons is of $5\times 10^4\,\msun$ but 
 this limit does not represent the minimum mass scale of resolution within our semi-analytical treatment of NSC formation. According to theoretical stdies, halos with mass
 less than  $M_{\rm h,min}$ do not fulfill the conditions for star formation and fragmentation
 at the redshifts explored \citep{barkana01,tegmark97}.

In our complex scheme (here simplified in its skeleton to outline only the key steps), virialized dark matter halos, forming at any  redshift, accrete baryonic gas in a fraction equal to $\Omega_{\rm baryon }/\Omega_{\rm matter}.$
The gas, initially in virial equilibrium, cools down at a rate 
computed according to the available cooling channel (either molecular or atomic cooling depending on the metallicity
of the gas in the halo) and condense into the centre of the halo.
The code distinguishes between metal free halos (those with metallicity $Z<Z_{\rm crit}$, where $Z_{\rm crit}\sim 10^{-4.87}\,Z_\odot$ is the critical metallicity for fragmentation), and halos enriched above $Z_{\rm crit}$ \citep{santoro06}. 

\begin{figure}
\includegraphics[scale=0.33]{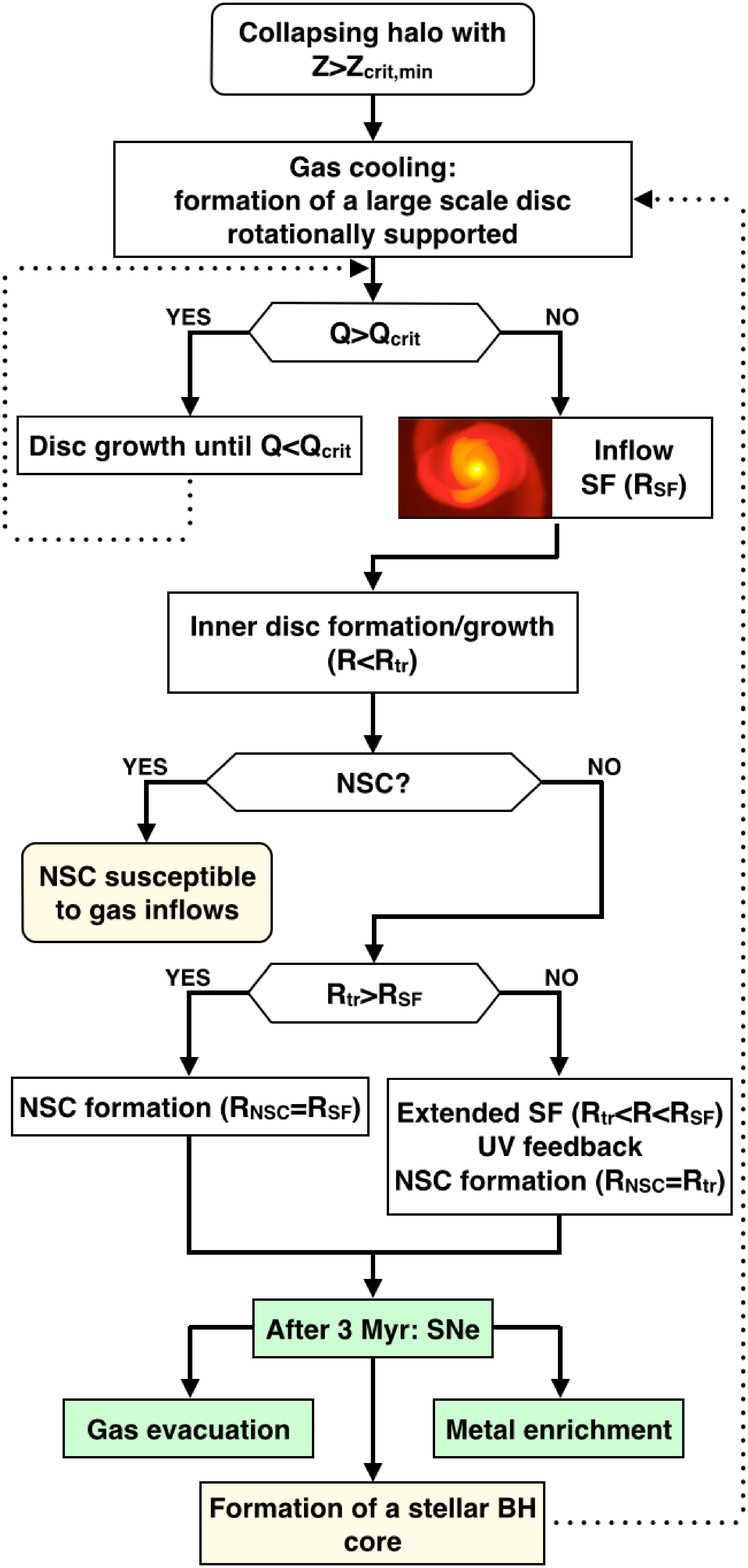}
\caption{\small{Flow chart of an isolated halo with $Z>Z_{\rm crit}$, starting from gas cooling until the formation of a NSC in the centre of the halo. The processes/events in the flow chart are described in \S ~3.}}
\label{fig:path}
\end{figure}

\begin{itemize}
\item
In metal free conditions, typical of the first collapsing halos, the halo virial temperature is $T_{\rm vir}<10^4$ K, and the only available cooling channel is that of molecular hydrogen (H$_2$).
If enough H$_2$ is present in the halo, gas can cool down to $T\sim 200$ K, reaching high enough densities to form a Pop III star. In the code only a single Pop III star is allowed to form in halos with $T_{\rm vir}\gsim 10^3$ K \citep{tegmark97} with a mass extracted from an top-heavy initial mass function extending from a minimum to a maximum mass of 10 and 300 $\msun$, respectively  (D10).

Pop III stars can produce a strong UV flux, able to affect the thermodynamics of gas inside the halo and in the neighbouring ones. In particular, the strong Lyman-Werner flux emitted is able to photo-dissociate H$_2$ molecules, thus quenching molecular cooling. Formation
of new zero metallicity stars is suppressed in those halos in which  the H$_2$ dissociation rate is higher than its formation rate. In massive
halos above a threshold mass self-shielding of the gas, in the denser regions, can however prevent the disruption of molecular hydrogen
\citep{machacek01,madau01} and this is accounted for in the simulation.
\footnote{Note that in presence of
a sufficiently high Lyman-Werner background, seed black holes can form via direct collapse of warm gas clouds in some halo \citep{agarwal12,latif13}.
This is a path which is not considered in this scheme.}

\item
In metal enriched halos (with typical virial temperatures $T_{\rm vir}>10^4$) gas cools down via both atomic and molecular cooling and collapses toward the halo centre. 
Assuming angular momentum conservation, the cool gas settles into a large-scale pre-galactic disc described as a
rotationally supported self-gravitating Mestel disc.  The disc keeps on growing in mass due to continuous infall of cool gas, and
the progressive increase in the gas mass causes the disc to become Toomre unstable.
Torques induced by self-gravity lead to a relatively fast redistribution of gas within the disc: the gas shocks and loses angular momentum thus sinking to the centre of the halo where it forms an inner disc. The inner disc develops a steeper profile than the outer unstable disc and will become
the site of NSC formation. 
As gas flows 
in the inner disc, the surface density in the outer Mestel disc decreases
until the Toomre parameter increases nearing  the critical value for stability ($Q_{\rm crit}\sim 2$,
in the simulations considered). At this time the central infall of gas stops.
The overall process  self-regulates to guarantee a condition of marginal stability in the outer disc. The mass routed in the central part 
and forming the inner disc then corresponds to the amount necessary for the outer disc to be marginally stable.
The final configuration will then be characterised by an outer disc with Mestel profile and an inner  disc with a steeper 
density profile (with surface density scaling as $R^{-5/3}.$
The physical parameters of the two nested discs are then calculated self consistently
with the transition between the outer and inner discs occurring at a transition radius $R_{\rm tr}$ (D10).

This holds true if fragmentation of the gas, conducive to star formation, does not take place overall in the outer disc. 
As reported in \citet{lodato07} fragmentation occurs when the gravitational induced stress into the disc is larger than a critical value.
  In the code, fragmentation is allowed where the cooling rate exceeds the adiabatic heating rate (D09). These two conditions determine the region where star formation (SF) sets in, which can be defined introducing a characteristic radius $R_{\rm SF}$.
If the outer disc grows strongly unstable, widespread star formation sets in consuming part of the gas that would otherwise flow into the inner disc. 
In this case $R_{\rm SF}>R_{\rm tr}$ and a NSC forms with a radius $R_{\rm cl}=R_{\rm tr}$ and a mass $M_{\rm cl}$ which is calculated self-consistently
considering that left at disposal after gas consumption.
By contrast, if the outer disc remains stable against fragmentation, star formation is triggered in the inner disc, and a NSC forms with a radius corresponding to the star formation radius, i.e. $R_{\rm cl}=R_{\rm SF}$ and a mass equal to the total mass 
of the inner disc turned into stars. A flow chart describing all the processes leading to the formation of a NSC, which has been described in this section, is reported in Figure \ref{fig:path}.

A parameter which determines the NSC mass, in the simulation, is $\eta$, controlling the extent of baryonic inflows from the outer to the inner disc.  
Since star formation in the outer disc reduces the amount of gas that flows in,  the net inflow rate, and thus the
mass of the NSC at birth, is computed as difference between the nominal inflow rate which an unstable outer disc would have
and that consumed in stars.
In our simulations we assume $\eta$, as defined in D10, equal to $1$.
As shown in D10, the net inflow rate and the mass of the NSC at birth turn out to have a weak dependence on $\eta$ (see figure 2 of D10) since 
with increasing $\eta$ more gas is consumed by extended star formation before it can reach the inner disc. The total gas mass that has flown inwards, at the end of the simulation, turned out to be slightly lower for higher values of $\eta$. 
The continuous inflow of gas is a key ingredient for the model under study, as the NSC after birth is invaded by repeated episodes of gas accretion, thus leading to the evolution that we will explore in the incoming sections. Hereafter, we hypothesise that gas onto the
NSC  acts to deepen the potential well without turning into stars. 

\item In our model, the physical consequences of SNae are included, for all channels, using motivated recipes, to account for gas depletion and metal pollution.
Three million years after NSC formation, it is assumed that the most massive stars explode as SNae, primarily leading to gas evacuation
at a rate correlated to the energy of the explosion. 
This affects the extent of gas inflows on the NSC at subsequent times, as 
star formation in the surroundings of the NSC is temporarily quenched.
Similarly, metal pollution from SNae is accounted for using a simple parametric prescription
which correlates the SNae rate with the rate of injection of metals.  A fraction $f_{\rm metal}=
1-M_{\rm sh}/M_{\rm sn}$, with $M_{\rm sh}$ the gas mass removed from the halo (see DV10) and $M_{\rm sn}$ the total mass released by the SN, is then retained in the halo.
We remark that in the explored range of metallicities, the masses of the NSC at birth turned out to depend weakly on the metal content (as shown in figure 3 of D10).

The very massive stars in the NSC are assumed to leave behind a relic population of stellar mass black holes which
form a core after mass segregating.
In this paper, the core represents the site of formation of the seed black hole after the NSC has been subjected to repeated gas inflows.

 \item 
In our cosmological scenario, halos undergo mergers, and the merger history of any halo can be obtained from Pinocchio.
In the code, merger prescriptions read: (a) the total, baryonic, stellar and metal masses of the new halo are computed as sum of the masses of the two progenitors; (b) if the mass ratio between the halos is less than 1/10 the spin parameter of the main progenitor is retained, otherwise a new value is computed following \citet{Bett07}; (c) the properties of the new pre-galactic disc are calculated taking into account the new mass and angular momentum, and then disc stability, eventual inflow and star formation are re-evaluated; 
(d) if one or both the halos host a seed black hole, different paths can be followed.
In a major merger the secondary black hole (if present) is able to reach the centre of the primary, while in a minor merger it is left wandering in the new halo \citep{Callegari11}. 
The code is able to describe the remnant at the end of the merger only, but it completely neglects all processes taking place during the merger events.
\item
NSCs start forming around redshift $z\sim 20$  when the universe is 180 Myrs old.  With the code we follow their evolution
down to redshift $z\sim 6$ when the universe is $\lsim$ 1 Gyr old. We then halt the simulations, since
the code does not implement any accretion history 
important at  lower redshifts \citep{merloni13}.

\end{itemize}
 
\section{Gas Inflows onto Nuclear star clusters}
In this section we first introduce the contraction parameter, describing the dynamical response of a NSC to an arbitrary inflow of gas, and later we
illustrate  the implementation of the new channel of black hole seed formation, in the cosmological simulation sketched in $\S$ 2.

\label{sec:gasdriven}

\subsection{Contraction parameter}
\label{sec:contrpar}
Consider the case of a NSC at the centre of a pre-galactic disc which is subjected to an inflow of gas, due to a 
perturbation inside the parent halo or due to a halo-halo merger, and let $M_{\rm gas}$ the total gas mass involved in the inflow. 
What is the kinematical response of the cluster to the inflow of gas?
If, in the simplest hypothesis, the gas contributes only to the deepening of the potential well, this causes a change in
the orbital velocity dispersion of the stars.
Assuming that stars in the cluster move on nearly circular orbits and that their angular momentum is conserved during the inflow event, we have
\begin{equation}
l_\star = m_\star\sqrt{GM(<r_0)r_0}= m_\star\sqrt{G(M(<r_0)+M_{\rm gas})r},
\end{equation}
where $r_0$ and $r$ are the stellar radii  before and after the inflow event, respectively, and  $M(<r_0)$ is the stellar mass within $r_0$. 
In response to the inflow, the new radius is 
\begin{equation}
r/r_0=M(<r_0)/[M(<r_0)+M_{\rm gas}].
\end{equation}
This is an over-simplifying assumption not only because stars do move on rosetta, eccentric orbits in star clusters but also because
no assumption has been made on how the gas is distributed inside the star's cluster, i.e. whether is more clustered toward the centre or
distributed over a much larger volume (assuming again for simplicity spherical symmetry).

In order to model different degrees of contraction for the cluster and redistribution of gas, we generalised this relation assuming a power law with a generic exponent $\xi$ for the entire cluster, which can be written as
\begin{equation}
R_{\rm cl}/R_{\rm cl, 0}=[M_{\rm cl, 0}/(M_{\rm cl,0}+M_{\rm gas})/]^{\xi},
\label{xi}
\end{equation}
where $R_{\rm cl}$ and $R_{\rm cl,0}$ are the cluster radii before and after the inflow event.
According to equation~(\ref{xi}), $\xi=1$.

\begin{figure}
\includegraphics[scale=0.3,angle=-90]{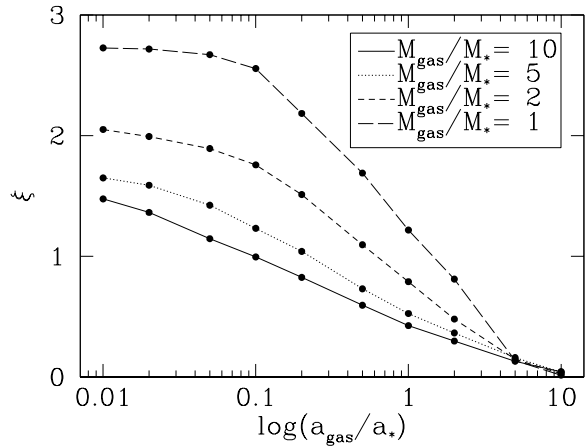}
\caption{\small{Contraction parameter $\xi$  from equation~(\ref{xi}) as a function of the ratio between the gas scale radius $a_{\rm gas}$ and the initial stellar scale radius $a_*$ for different inflow masses.}}
\label{fig:xi}
\end{figure}

The running value of the parameter $\xi$ can be
estimated more accurately, exploring as toy model a spherical inflow, of given mass $M_{\rm gas}$, in a star cluster described by a Plummer
potential of scale radius $a_*$ initially. 
If during a spherical inflow, the angular momentum per unit mass of individual stars is conserved, 
then $\xi$ varies between $\sim 0$ and $\sim 3$, as illustrated in  Figure~\ref{fig:xi}.  This figure is obtained 
under the assumption  that also the gas after the inflow settles into virial equilibrium following a Plummer profile
with scale radius $a_{\rm gas}$.  
In Figure \ref{fig:xi} we plot $\xi$ as a function of $a_{\rm gas}$ for different values of the  mass inflow  $M_{\rm gas}$. Varying $a_{\rm gas}$ is a way to mimic
different cooling prescriptions for the gas which can contract down to a very small radius  (in the absence of 
a modeling of the thermodynamical behaviour of the  gas). As expected, the largest contraction 
occurs, for a given $M_{\rm gas}$, in the limit of  $a_{\rm gas}\to 0$ and is larger  the larger $M_{\rm gas}$ is.  
For a Plummer sphere, $\xi$   saturates to a value equal
to $\ln(2^{3/2})=2.8$  in the limit in which the whole mass $M_{\rm gas}$ has collapsed into a point.
\footnote{The analytical estimates of  $\xi$ shown in Figure~\ref{fig:xi} match well with 
those obtained calculating the response of a star cluster to an external inflow, using a Monte Carlo
integrator for the true dynamics of the stars (our stellar black holes)  in a Plummer sphere (Galanti et al. 2014, in preparation).}

From this toy model we notice that cluster contraction depends on how the gas concentrates within the cluster.
Since the inflow of mass on the pre-existing star cluster can occur at any time according to the evolution of the halo's cosmic environment, and 
can vary from episode to episode, the contraction parameter is very susceptible to conditions 
that allow for a wide range of values. To bracket uncertainties we thus explore NSCs with inflows varying 
the contraction parameter $\xi$ between a minimum value of 0.5 and a maximum of 3.
The contraction of the star cluster due to the gas inflows also leads to an increase of the velocity dispersion of the stars which is computed according to the virial theorem, once the new radius and new total cluster's mass are known. 
 
\subsection{Modelling the Gas Induced black hole Runaway process in NSC with inflows}
\label{sec:GIRM}

\begin{figure}
\includegraphics[scale=0.33]{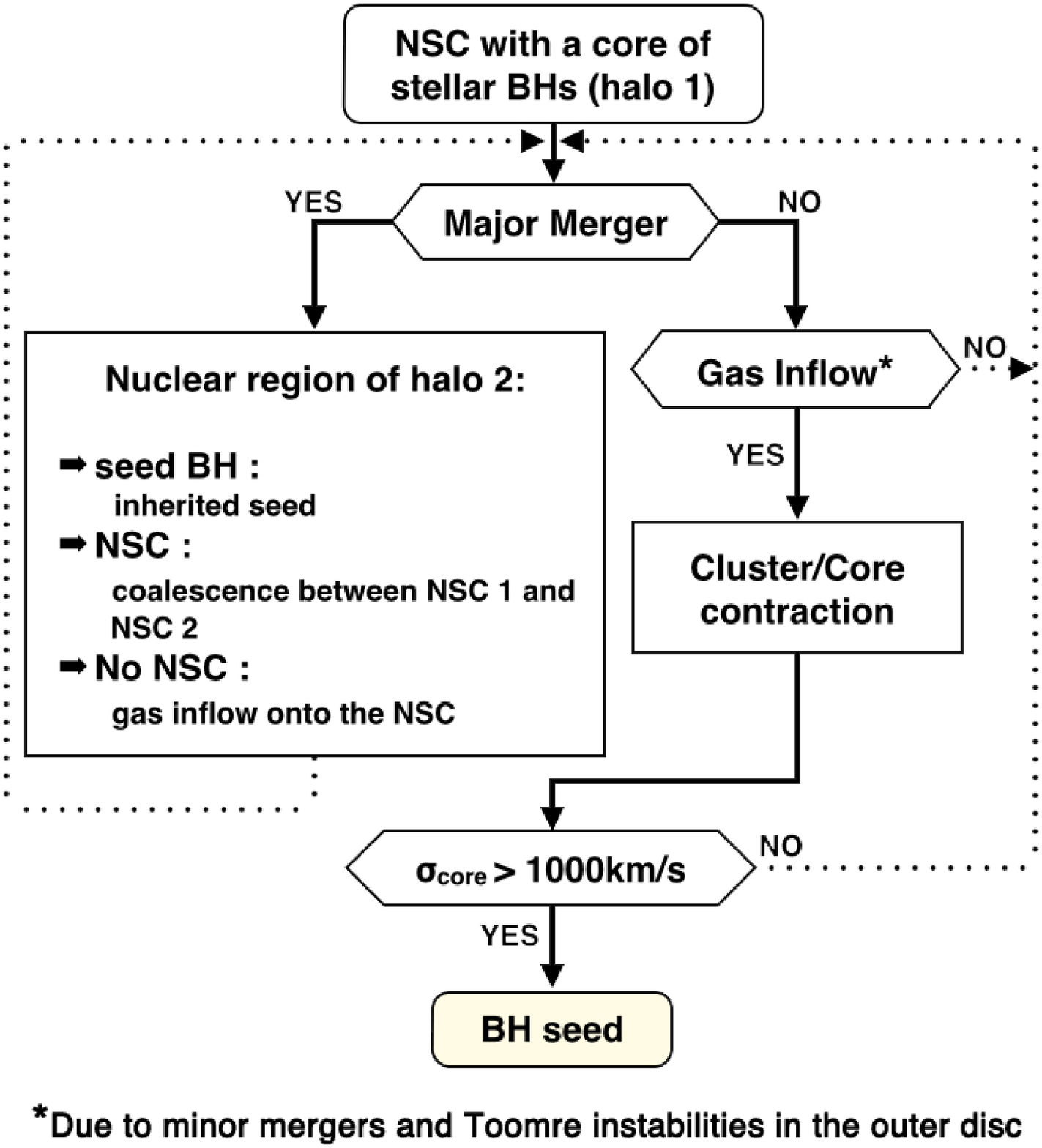}
\caption{\small{Flow chart reporting the NSC contraction process leading to the formation of a single seed BH from the merger among stellar mass black holes.}}
\label{fig:NSCflow}
\end{figure}

Let $M_{\rm cl,0}$, $R_{\rm cl, 0}$ and $\sigma_{\rm cl}$ be the mass, radius and 1D velocity dispersion of the star cluster at the time
of its formation.  
Stellar mass black holes are assumed to form after the first three Myrs and to mass segregate
shortly after within a core radius  $R_{\rm BH-core,0}=0.1R_{\rm cl,0}$.
The typical mass segregation timescale can be computed as
\begin{equation}
t_{\rm segr} \approx \frac{\langle m_\star\rangle}{m_{\rm BH}}\frac{N}{8\log N}\frac{R_{\rm cl,0}}{v_{\star,0}},
\end{equation} where $\langle m_\star\rangle$ is the mean stellar mass, $m_{\rm BH} $ the 
mean black hole mass, $N$  the number of stars and black holes in the cluster, and $v_{\star,0}$ is the stellar typical velocity which can be approximated with the 1D velocity dispersion $\sigma_{\rm cl,0}$ of the cluster.
If we consider a typical NSC initial mass of $10^4\, \rm M_\odot$, with radius $R_{\rm cl,0} = 1$ pc, mean stellar mass of  
 $\langle m_\star\rangle \simeq 1\rm\, M_\odot$ ,and  $m_{\rm BH}\sim 10\rm\, M_\odot$, the mass segregation timescale is $\sim 4 $ Myr.
We thus assume that a core of stellar black holes forms 4 Myr after the NSC formation. 
We further define $M_{\rm BH-core,0}$ as the mass in stellar black holes present in the cluster core, and $\sigma_{\rm BH-core,0}$ the black hole velocity dispersion: we assume  $M_{\rm BH-core,0}=1.5\times 10^{-3}M_{\rm cl,0}$ \footnote{This relation is derived from $M_{\rm core,0}=m_{\rm BH}N_{\rm BH}$, where $N_{\rm BH}=\alpha N_\star$ with $\alpha=0.0015$, and $M_{\rm cl,0}=\langle m_\star \rangle N_\star+m_{\rm BH}N_{\rm BH}$ (see DMB11). }  and $\sigma_{\rm BH-core}=26\sigma_{\rm cl}$, where 26 corresponds to virial value $(GM_{\rm BH}/R_{\rm BH-core,0})^{1/2}$.\\
Figure \ref{fig:NSCflow} shows a schematic picture of the NSC contraction process driven by gas inflow, ending with the formation of a seed black holes.

\noindent
GIRM is implemented in the code as follows:
\begin{itemize}
\item
Given a mass inflow $M_{\rm gas}$ onto the NSC, the radius $R_{\rm cl}$ is updated according to equation~\ref{xi}, 
where the exponent $\xi$ is allowed to vary between 0.5 and 3, as explained in section \ref{sec:contrpar}. 
The new velocity dispersion $\sigma_{\rm cl}$ is computed
from the virial theorem, considering as mass the sum of the cluster and gas mass.
Gas is allowed to settle into a
disc of size $R_{\rm tr}$, as a purely radial inflow would require an unphysical transport/cancellation of angular momentum.
The gas disc has a surface density profile of the form  $\Sigma_{\rm gas}=\Sigma_0(R/R_{\rm tr})^{-5/3}$ 
with $\Sigma_0=
M_{\rm gas}/2\pi R_{\rm tr}^2$.

\item 
We compute the gas mass $M_{\rm gas,core}$ enclosed in $R_{\rm BH-core}$ from the above relation for
the disc and update the stellar core radius $R_{\rm BH-core}$ 
according to equation (\ref{xi}).
The 
enhancement of the core 
velocity dispersion $\sigma_{\rm BH,core}$ is computed from the virial relation using as mass the sum of $M_{\rm BH-core,0}$  
and $M_{\rm gas,core}$.
If, in response to an inflow, $\sigma_{\rm BH-core}$ exceeds 1000 $\kms$
we assume that the whole core of stellar black holes  experiences  GIRM, which leads to the formation of  a seed of mass 
equal to $M_{\rm BH-core,0}$.
The choice of 1000 $\kms$ is motivated by the study of DBM11, representing the threshold for GIRM.
In the DBM11 scenario, 
hard binaries coalesce before experiencing single-binary encounters and, at the same time, are retained
in the core as kicks from gravitational recoil are not large enough to allow black holes to escape from the cluster
(having an escape speed $\sim 4000 \, \kms$).

\item 
In presence of halo-halo mergers, NSCs are modelled using simple recipes which account for the 
additional inflows of gas (\citeauthor{mihos96}). In the case of major mergers (those with mass ratio larger than 1:10)
we follow these prescriptions:  (a) if only one halo hosts a NSC prior to the merger, the gas of the companion halo 
is added to the new halo causing an inflow in the first, and the cluster and core parameters are
updated considering a rapid infall. The new radius and mass of the star cluster with its embedded black hole 
is computed using the contraction law;
(b) if both halos hosted a NSC, we merged the two  into a single new cluster, and update the cluster properties following the prescriptions in \citet{ciotti07}.
Assuming that during the merger no mass is lost, energy conservation allows to estimate the virial radius $R_{12}$ of the new cluster core of black holes 
as
\begin{equation}
\frac{1}{R_{12}}=\left(\frac{M_{\rm BH,1}+M_{\rm g,1}}{M_{\rm BH,12}+M_{\rm g,12}}\right)^2\frac{1}{R_1}+\left(\frac{M_{\rm BH,2}+M_{\rm g,2}}{M_{\rm BH,12}+M_{\rm g,12}}\right)^2\frac{1}{R_2},
\end{equation}
where $M_{\rm BH,1}$ and $M_{\rm BH,2}$ are the masses of the black hole cores of the progenitor clusters, $M_{\rm g,1}$ and $M_{\rm g,2}$ the gas masses in the cores, $M_{\rm BH,12}$ and $M_{\rm g,12}$  the total mass in black holes and gas of the new star clusters, respectively, and $R_1$ and $R_2$ are the virial radii of the progenitor clusters;
(c) if one halo had already a seed black hole with a mass above 260 M$_\odot$ relic of a Pop III star, this black hole
is incorporated in the resulting halo as a new seed;
(d) if both halos hosted a seed black hole we assumed the instantaneous coalescence of the two black holes into a more massive seed.

\item
In the case of minor mergers, NSCs are modelled assuming that the only change in their properties is due to an additional gas inflow. Black holes and clusters from the smaller halo are let wandering in the outer region of the resulting galaxy, and are not allowed to sink into the nuclear region \citep{Callegari11}.

\item
 We further explored, as ancillary study, NSCs with velocity dispersions in the interval between $40$ and $100\,\kms$. 
These NSCs have no relation with the GIRM scenario, but their presence can be easily tracked, within the cosmological scheme explored here. 
Their relatively high velocity dispersion (between $40$ and  $100 \,\kms$)  can be natal or more likely is acquired following a gas inflow. The request
on the velocity enhancement here is less severe than in GIRM. 
This will enable us to investigate on the likelihood of an alternative scenario by \citet{miller12}. In this new context, 
NSCs with velocities dispersions in this selected range, can grow a black hole seed from a stellar mass black hole (or binary black hole).
Stellar black holes, relic of massive stars, have two possible fates in these NSCs: they are either all ejected or a single (or binary) remains.  The black hole (binary) which avoided ejection, can later grow via tidal captures.  Dynamical arguments by \citet{miller12} seem to support this picture. 
We thus aim to calculate the number of NSCs with velocity dispersion between $40$ and $100\,\kms$, resulting either from the direct collapse of an unstable disc, or after a phase of gas inflow into a pre-existing cluster.
\end{itemize}

\section{Results}
\label{sec:results}
\subsection{Seed black holes from GIRM in nuclear star clusters}

We ran a suite of simulations, for three different values  of $\xi=$0.5, 1 and 3 for the contraction parameter.
The cosmological box explored has a size of 10 Mpc and has been evolved from redshift $z\sim 40$ 
down to redshift $z\sim 6$, to test if the GIRM channel is able to produce a population of seed black holes large enough to explain the most massive black holes found in the Universe at redshift up to $\sim 7$. We halt integration at $z\sim 6$ 
as our code does not include any process of accretion that can become important below this redshift.

In the simulations  we considered the Pop III and GIRM channels.
They are expected to contribute to the early generations of seed black holes at different cosmic epochs, as the formation of NSCs in pre-galactic discs (a necessary condition for the GIRM to operate) requires some degree of metal pollution of the intergalactic medium from the explosions of the massive Pop III stars. We remark that the GIRM channel does not pose any constraint on the level of metallicity of the parent halo, and hence on the time of formation, provided Pop III stars have enhanced the metallicity above a  threshold $Z_{\rm crit}\sim 10^{-4.87}\,Z_\odot$ (see D10-D12).  The GIRM path requires the formation of
ordinary star clusters and thus a population of stellar mass black holes from the core-collapse of the massive stars.

\begin{figure}
\centering
\includegraphics[scale=0.3,angle=-90]{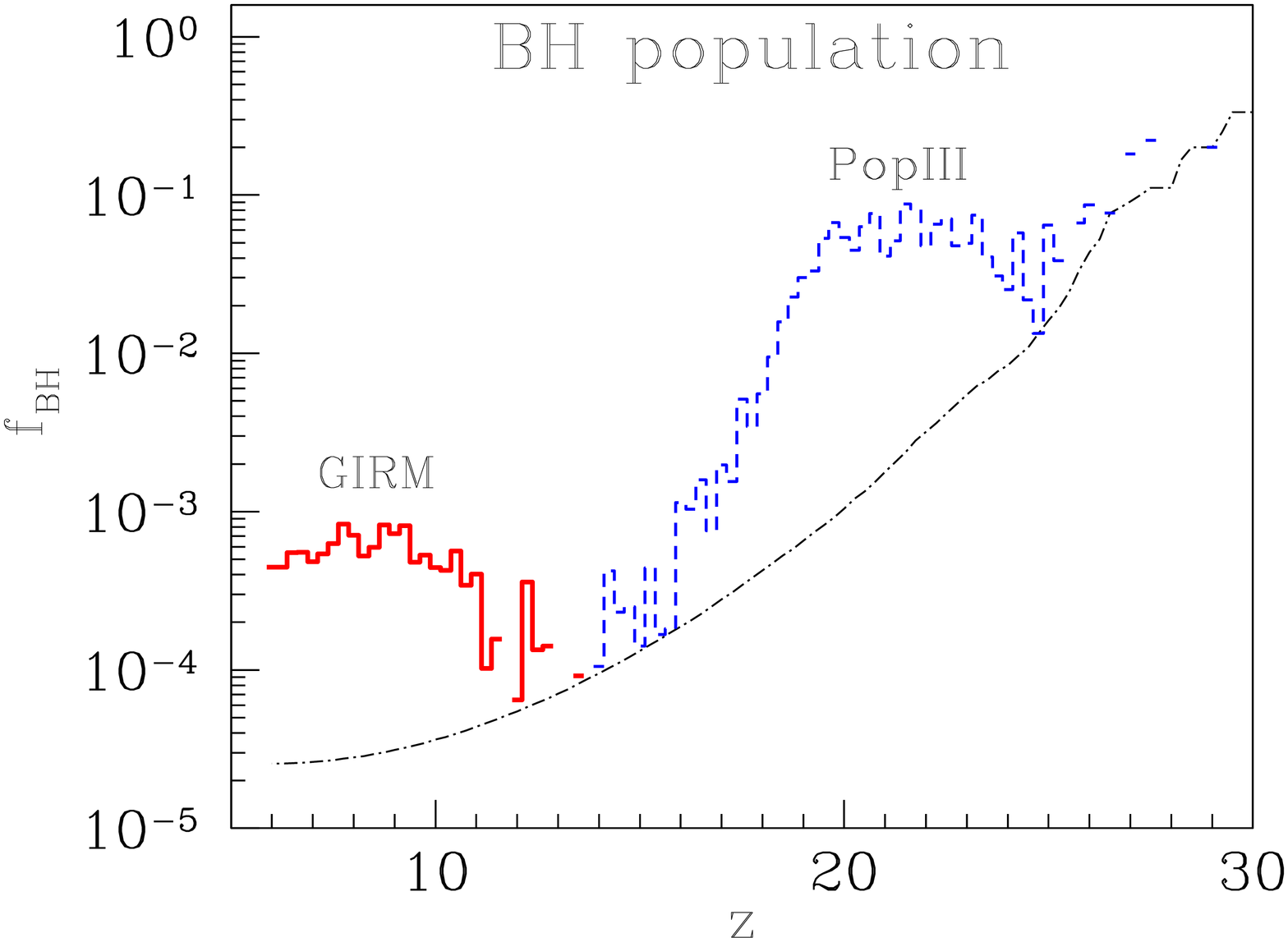}
\caption{{Fraction $f_{\rm BH}$ of halos hosting a newly formed seed black hole 
as a function of redshift, from redshift  $z=30$ down to $z=$6.  Blue line refers to the Pop III channel; red solid line refers to the fraction of black holes with mass above $260\,\msun$ resulting from the GIRM channel, for a contraction parameter $\xi=0.5$ The black dash-dotted line is the simulation resolution.}}
\label{fig:fbh0.5}
\end{figure}

\begin{figure}
\centering
\includegraphics[scale=0.3,angle=-90]{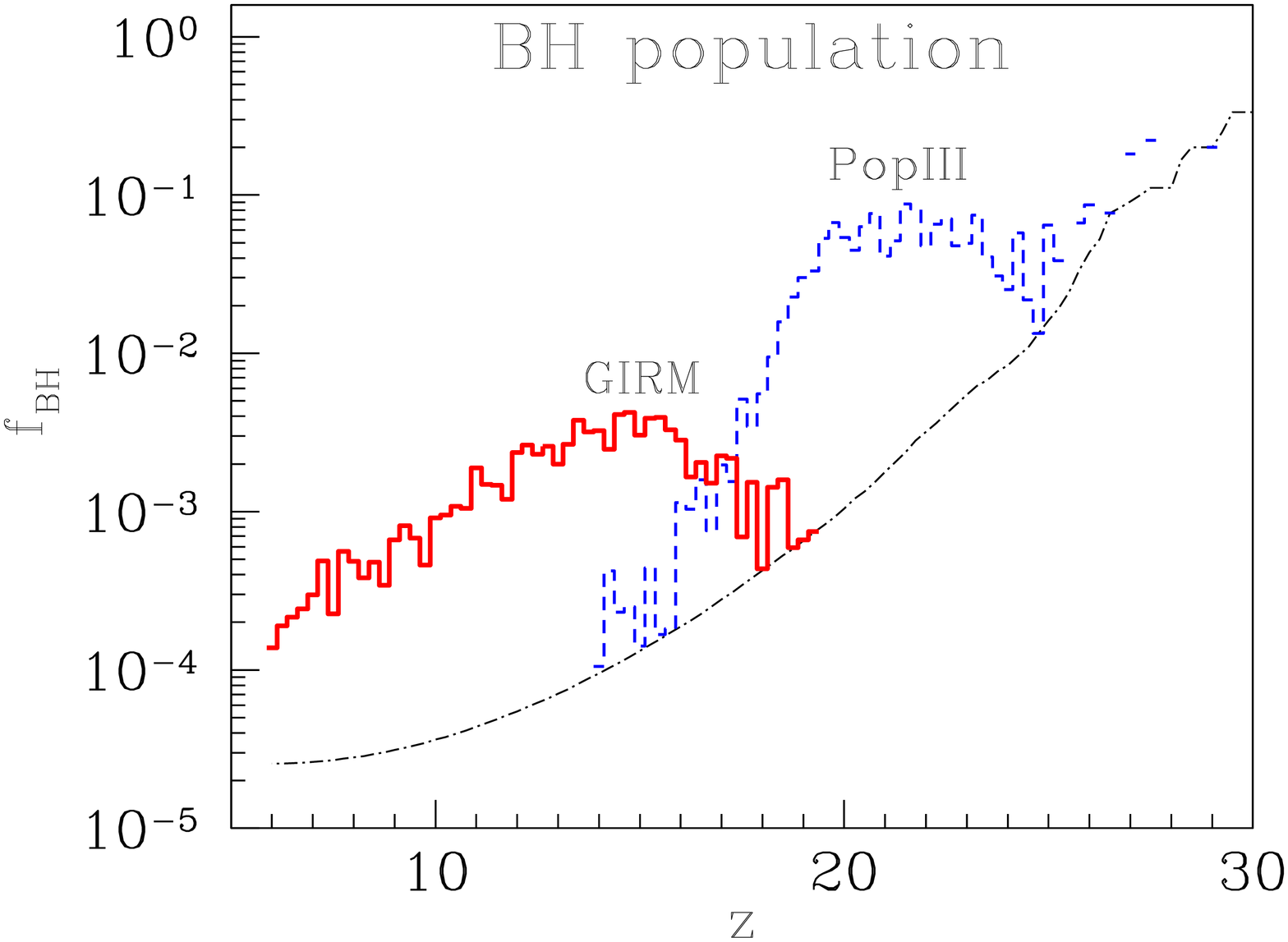}
\caption{{ 
Fraction $f_{\rm BH}$ of halos hosting a newly formed seed black hole 
as a function of redshift, from redshift  $z=30$ down to $z=$6.
Blue line refers to the Pop III channel; red solid line refers to the fraction of black holes with mass above $260\,\msun$ resulting from the GIRM channel for a contraction parameter $\xi=3$ The black dash-dotted line is the simulation resolution.}}
\label{fig:fbh3.0}
\end{figure}

Figures \ref{fig:fbh0.5} and \ref{fig:fbh3.0} show the number of halos relative to the total
 $f_{\rm BH}=N_{\rm BH,z}/N_{\rm halo,z}$ which host a newly formed black hole
 at any given redshift.\footnote{The case for $\xi=1$ is intermediate between Figures \ref{fig:fbh0.5} and \ref{fig:fbh3.0} and is not reported for seek of simplicity.} 
 The simulation resolution reported in the plots is computed as $1/N_{\rm halos,z}$.
 
\begin{figure}
\includegraphics[scale=0.3,angle=-90]{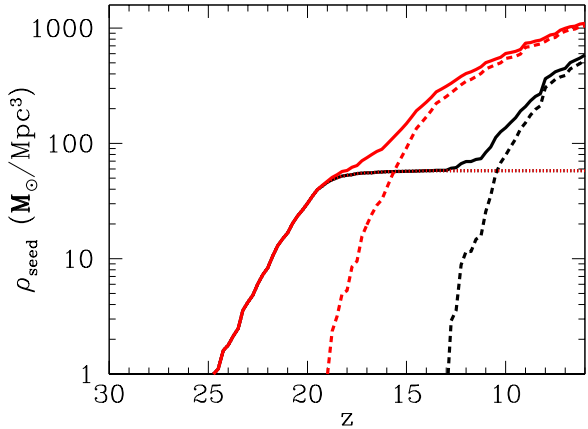}
\caption{\small{ Black hole mass density versus redshift for the two channles. The dotted line 
refers to the Pop III path. The dot-dahed lines refer to the GIRM channel for 
 for $\xi=0.5$ (right)  and $\xi=3$ (left). The total seed mass densities are also plotted with the thin 
 solid  lines,  for $\xi=0.5$ (right) and the solid one for $\xi=3$ (left).}}
\label{fig:rhoseed}
\end{figure}

There exists an earlier epoch of black hole seed formation from Pop III, around $z\sim 20$, and 
a second,  from the GIRM channel around $z\sim 8,10$ and 15 according to the values of $\xi$ (as
given in Table~\ref{tab:NbhCh}), with an overlap with the Pop III channel for $\xi=3$.  A more rapid contraction of the star cluster
in response to the gas inflows (i.e. for $\xi=3$) makes black hole formation to occur earlier and to produce a larger number of seeds.
Later, the observed decline mirrors the bottom up hierarchical build up of the halos and of their embedded star clusters: 
less massive halos form first and require lower gas inflows to produce a seed, while heavier halos require a greater amount of gas
and are less in number. 
In the case of $\xi=0.5,$ black hole seed formation occurs at later epochs as there is the need of a 
larger inflow of gas to trigger the collapse of
the black hole star clusters.  The decline in the Pop III channel is instead due to 
metal pollution from the first Pop III stars which quenches the formation of
black hole seeds from pristine gas.  
Table~\ref{tab:NbhCh}  shows that the number of black hole seeds in the cosmological box increases with increasing $\xi$, and that the channel is competitive to that from Pop III. The fraction $f_{\rm BH}$ is larger for the Pop III channel, but 
with cosmic time, the absolute number of sufficiently massive dark matter halos increases so that 
the absolute number of natal black hole halos is larger as $z$ decreases. 
The evolution of the relative contribution to the black hole seed population due to Pop III and GIRM channels is depicted  in Figure \ref{fig:rhoseed}, where we report  the total seed comoving mass density as a function of redshift. 
In the plot the different lines correspond to the Pop III channel and GIRM channels for different values of $\xi$. In addition, also the total seed co-moving mass density has been represented. 

As in D10, black hole seeds from PopIII stars have a typical small mass ($\sim 70 \,\msun $), while
 seeds from the GIRM channel have a larger mass, of about $\gsim 1000\,\msun,$ as the entire cluster of stellar black holes
 is assumed to undergo runaway merger, after a major inflow of gas. 
The mean mass of black holes formed via the GIRM channel decreases with increasing $\xi$ and this can be explained with the larger fraction of 
NSCs with smaller masses able to core collapse before $z=6$. In general, to form a seed the required inflow of gas
is at least comparable with the mass of the target star cluster.

\begin{table}
\centering
\begin{tabular}{l|c|c|c|c}
\hline
$\xi$ & $\#_{\rm seed}$ &  $\langle M/\msun\rangle$ & $z_{\rm peak}$ \\
\hline\hline
0.5 & 405 & 1301.1& 7.75\\
1.0 & 637 & 1173.4 & 10.0 \\
3.0 & 1054 & 987.5 & 14.75\\
\hline
\hline
PopIII  & 822 & 70.5 & 27.5\\
\hline
\end{tabular}
\caption{{\small The number of black hole seeds formed from the GIRM channel (upper panel) and from Pop III (lower panel);  the mean black hole seed mass and the redshift where their formation peaks, considering the three adopted values of $\xi$.}}
\label{tab:NbhCh}
\end{table} 

\begin{figure}
\centering
\includegraphics[scale=0.3,angle=0]{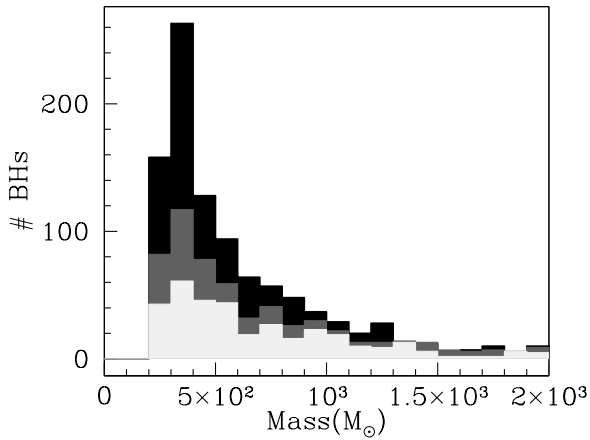}
\caption{{Mass distribution of seed black holes formed via the GIRM channel, for $\xi=0.5$, $\xi=1$ and $\xi=3$.
Black histogram corresponds to $\xi=3$, grey to $\xi=1$, and  beige to $\xi=0.5$.  Black hole seeds have a typical mass of about $350\,\msun$, while the mean mass, depends on the $\xi$,
and is given in Table~\ref{tab:NbhCh}}}
\label{fig:mass}
\end{figure}

\begin{figure}
\centering
\includegraphics[scale=0.3,angle=0]{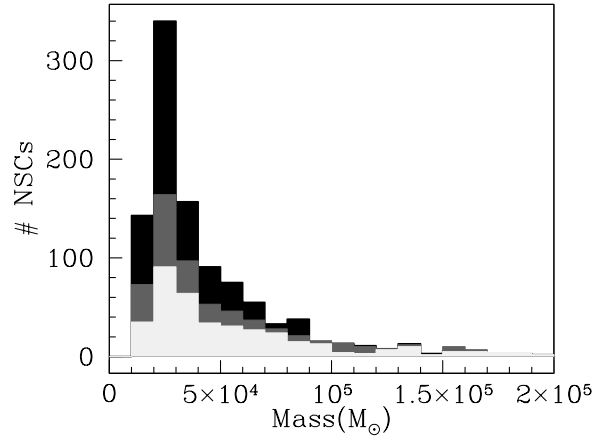}
\caption{{\small Mass distribution of the NSCs hosting seed black holes formed  via the GIRM channel, for $\xi=0.5,1$, and 3. Black histogram corresponds to $\xi=3$, grey to $\xi=1,$ and  beige to $\xi=0.5$.The largest fraction of
nuclear star clusters have a typical mass, prior to the gas inflow,  of about $3.5\cdot 10^4 \,\msun$.}}
\label{fig:mcl}
\end{figure}

Figure \ref{fig:mass} shows the black hole seed mass distribution, for the GIRM channel, for the three values of $\xi$, 0.5, 1, and 3. As already mentioned, the number of black hole seeds increases with increasing $\xi$. Despite the different mean mass, the peak of the mass distribution is around the same value of $350\,\msun$, mirroring the shape of the mass distribution of the parent NSCs housing the seeds, depicted in Figure~\ref{fig:mcl}.
The mass of the parent cluster is measured prior to the mass inflow that led to the core collapse. 

\begin{figure}
\centering
\includegraphics[scale=0.3,angle=-90]{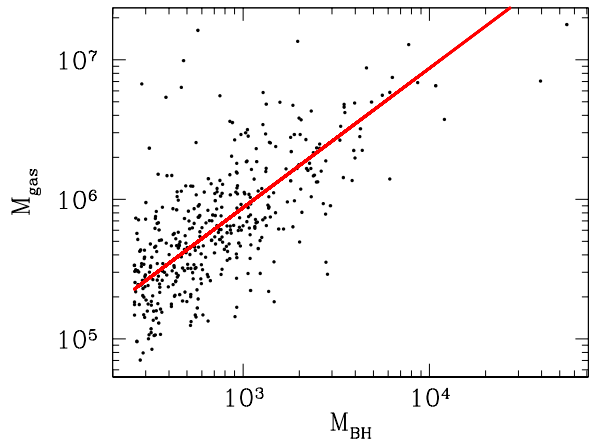}
\caption{{\small Gas mass resulting from inflows as a function of the seed black hole mass for the GIRM channel with $\xi=0.5$. The red solid line corresponds to the linear fit obtained assuming a zero black hole mass for  $M_{\rm gas}=0$.}}
\label{fig:mcl05}
\end{figure}

\begin{figure}
\centering
\includegraphics[scale=0.3,angle=-90]{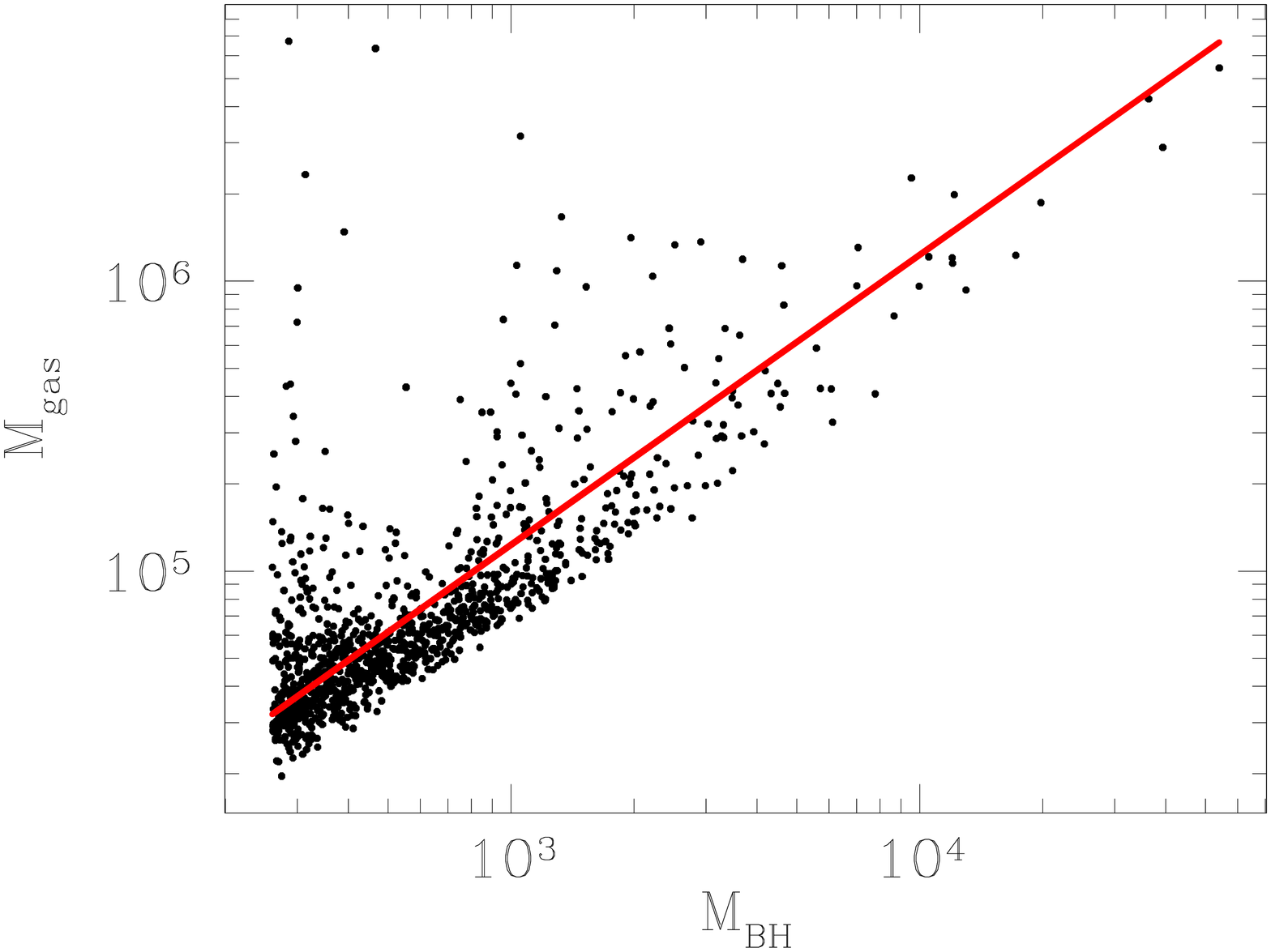}
\caption{{\small Gas mass resulting from inflows as a function of the seed black hole mass for the GIRM channel with $\xi=3$. The red solid line corresponds to the linear fit obtained as in Figure~\ref{fig:mcl05}.}}
\label{fig:mcl3}
\end{figure}

In addition, Figures \ref{fig:mcl05} and \ref{fig:mcl3} give the total mass of the inflow (due to repeated accretion events) within the transition radius $R_{\rm tr}$ as a function of the  black hole seed mass, evaluated when the velocity dispersion in the core reaches $\sigma\gsim 1000\,\kms$.
The red lines represent the best fit obtained assuming that $M_{\rm gas} \propto M_{\rm BH}$.
When $\xi=0.5$ the best line gives  $M_{\rm gas} \simeq 123 M_{\rm BH}$, while for $\xi=3$ we obtained $M_{\rm gas} \simeq 870 M_{\rm BH}$.
These results show that the total inflow needed to drive the NSC core to $\sigma \gsim 1000\,\kms$ varies from a minimum corresponding to roughly the NSC mass up to values as large as $10-50$ times the NSC mass.

\begin{figure}
\centering
\includegraphics[scale=0.3,angle=-90]{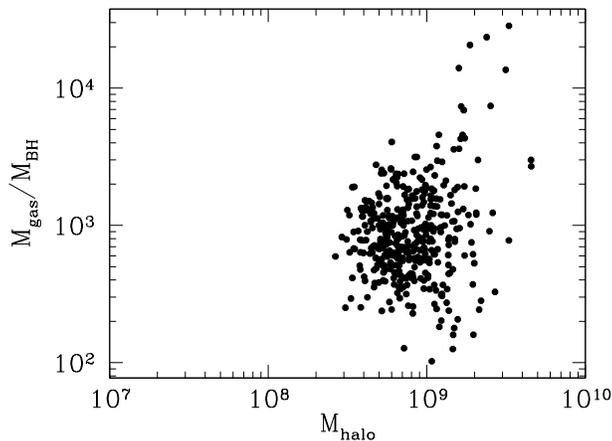}
\caption{{\small  Gas mass to seed black hole mass ratio resulting from inflows as a function of the host halo mass for the GIRM channel with $\xi=0.5$. The ratio is almost independent of the host halo mass, with the exception of the very large ones (where the inflow could be quite larger than that needed to GIRM) and it is almost constant with a value with an order of $10^3$.}}
\label{fig:infhalo05}
\end{figure}

\begin{figure}
\centering
\includegraphics[scale=0.3,angle=-90]{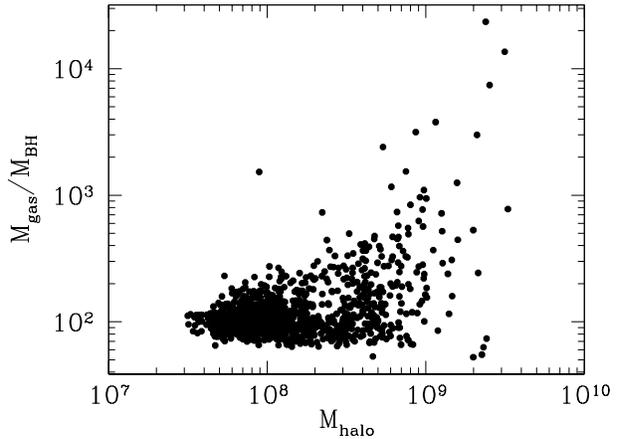}
\caption{{\small Gas mass to seed black hole mass ratio resulting from inflows as a function of the host halo mass for the GIRM channel with $\xi=3$. Also in this case the ratio weakly depends on the host halo mass, except for the very large ones (same as in Figure \ref{fig:infhalo05}).}}
\label{fig:infhalo3}
\end{figure}

As the extent of a gas inflow is related to the halo gas fraction and to mergers, we also expect a correlation between the total inflow needed for the GIRM to occur and the host halo mass.
Combining these two relations we find that the ratio between the gas inflow mass and the seed black hole mass is almost independent of the host halo mass, as reported in Figures \ref{fig:infhalo05} and \ref{fig:infhalo3}. 
In the case of $\xi=0.5$ only the most massive halos ($> 3\times 10^8\,\msun$) can host a seed black hole from GIRM, while for
a more efficient contraction ($\xi=3$) seeds are hosted also in less massive halos. 
At very large halo masses the ratio varies also of two orders of magnitude, and the largest values are probably due to halos with very large inflows, overcoming the necessary threshold for GIRM to occur.

\begin{figure}
\centering
\includegraphics[scale=0.3,angle=-90]{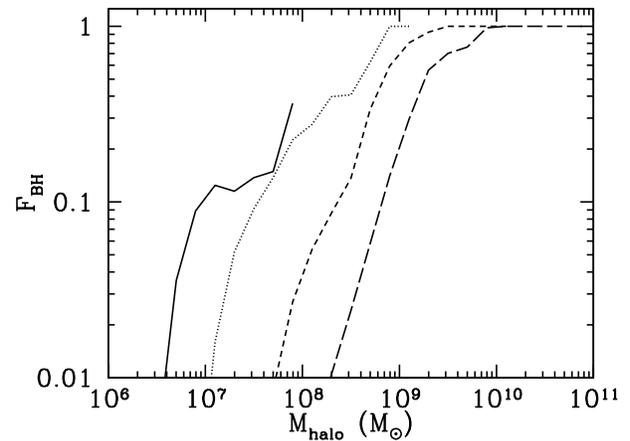}
\caption{{\small Occupation fraction of seed black holes versus halo mass assuming $\xi=0.5$. From left to right the lines show values at redshift $z=20,15,10,6$, respectively.}}
\label{fig:fmhalo05}
\end{figure}

\begin{figure}
\centering
\includegraphics[scale=0.3,angle=-90]{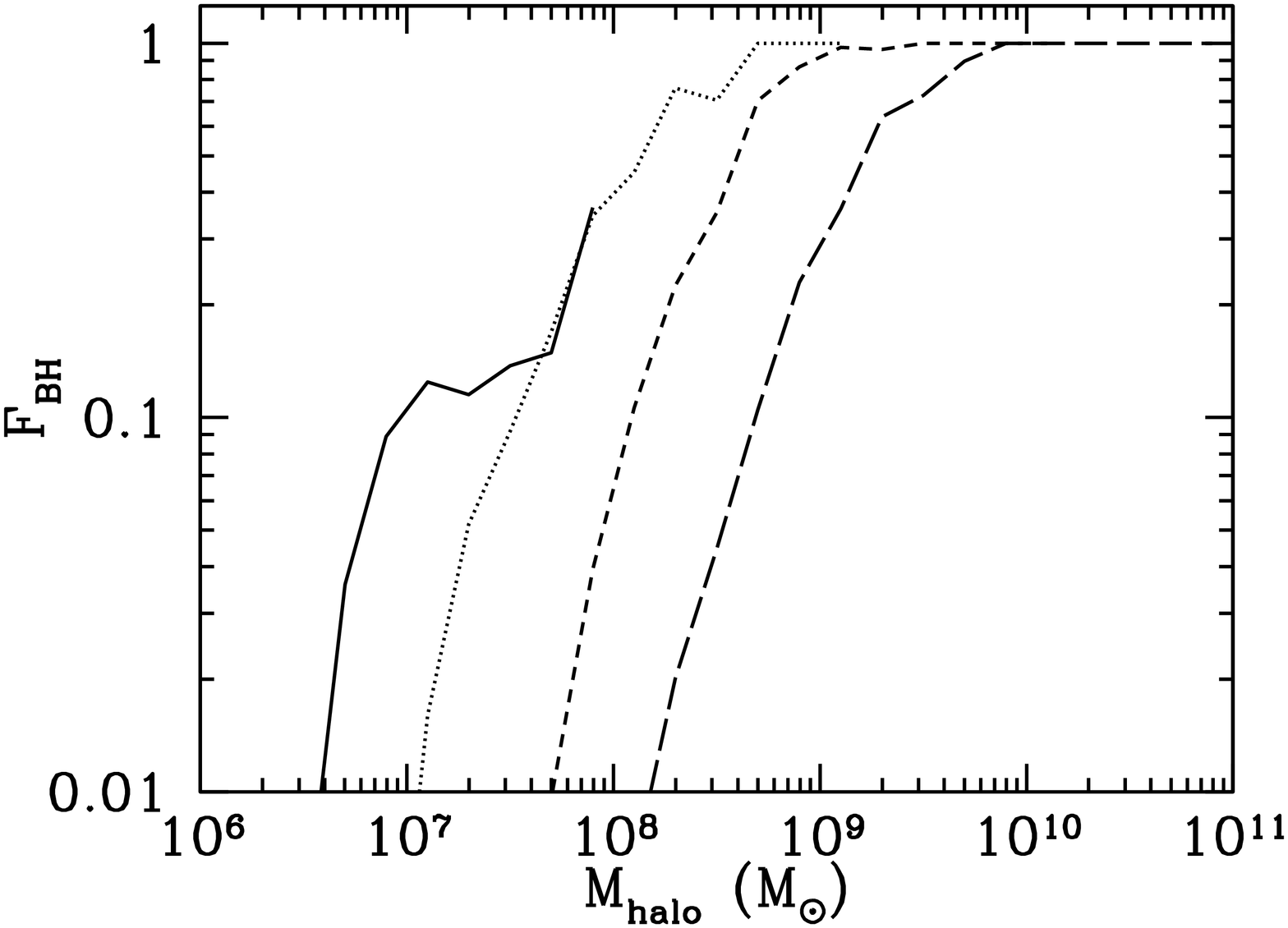}
\caption{{\small Occupation fraction of seed black holes versus halo mass assuming $\xi=3$. From left to right the lines show values at redshift $z=20,15,10,6$, respectively.}}
\label{fig:fmhalo3}
\end{figure}

Moreover, we find that the occupation fraction of black holes $F_{\rm BH}$  (formed via both Pop III and GIRM channels) is a function of redshift, and it strongly depends on the host halo mass.
Figures \ref{fig:fmhalo05} and \ref{fig:fmhalo3} show that at redshift $z=6$ halos with masses larger than $8-9 \times 10^9\rm\, M_\odot$ 
host a black hole, and halos with masses lower than $2 \times 10^8\rm\, M_\odot$ do not at all.
Note also the weak dependency on  $\xi$ of $F_{\rm BH}$.
For higher redshift, all the lines shift to lower masses, according to our cosmological model.

\subsection{Stellar black holes in nuclear star clusters}
\begin{table}
\centering
\begin{tabular}{l|c|c|c|c}
\hline
$\xi$ & $\rm max\{\#_{\rm seed}\}$ &    $<M_{\rm cl}/\msun>$ \\
\hline\hline
0.5 & 4003  & $5.83\cdot 10^4$\\
1.0 & 4421 & $5.22\cdot 10^4$\\
3.0 & 4814  & $4.90\cdot 10^4$\\
\hline
\end{tabular}
\caption{{\small The number of NSC candidates for core collapse following a gas inflow or a merger, leaving a single stellar mass black hole after 
ejection of the black holes in the core, as discussed in  Section 6, for the three different values of $\xi$.}}
\label{tab:Nbhsg40}
\end{table}

As anticipated in Section~\ref {sec:DMhalo}, NSCs with velocity dispersions in the interval  $40\,\kms$ and $100 \,\kms$ may be able to retain
a black hole or a binary black hole of stellar mass, as suggested in \citet{miller12}. These clusters have 
gravitational potential wells deep enough, 
and two-body relaxation times short enough for mass segregation to lead to formation of a core of stellar black holes. Having these clusters
high escape speeds, black hole ejection by three-body encounters 
or gravitational wave recoil is incomplete and either a single stellar black hole or a black hole binary remains
in the cluster which may later grow by capturing stars. 
Here, we briefly touch upon this model, by counting,  in this sub-section, the number of NSCs forming inside the
pre-galactic discs  that have their velocity dispersion above $40\,\kms$, or rise it due to a cosmological mass inflow or a merger.
The results are given in Table~\ref{tab:Nbhsg40} and can be considered as an upper limit to the stellar black hole population retained in these systems.    
The calculation
indicates that at the centres of pre-galactic discs there might be the conditions to build  star clusters 
 in which an intermediate mass black hole of stellar origin can form, and this could be a further channel of formation of seeds.
 From a comparison between Table~1 and Table~2, we note that the number of NSCs where this scenario may happen is 4-10 higher than in the standard GIRM.

\section{Conclusions}
\label{sec:conclusions}
In this paper, we explored the formation of black hole seeds of $\lesssim 10^3\,\msun$, resulting from the 
dynamical collapse of  stellar black holes in NSCs undergoing major gas inflows, at the centre of 
pre-galactic discs forming at very high redshifts.

We explored the properties of this population of forming seeds combining the clustering of dark matter halos 
extracted from``Pinocchio''
with a semi-analytical model which incorporates the physics of the inflows in a simplified form. Inflows onto the central star clusters
are triggered either by instabilities in the evolving pre-galactic disc and/or from mergers. 
Our aim was at comparing the efficiency of the GIRM channel with the already explored channels  from Pop III 
(here reported as a reference)  and from the collapse of supra-massive stars 
in young dense star clusters.
 
Our channel is active around redshift $z\sim 10-15$ as soon as the first NSCs 
form in the dark matter halos.  
Contrary to the two channels mentioned above, the GIRM is insensitive to the metallicity of the 
environment. The GIRM, by hypothesis, can occur in any  star cluster 
that formed a core of ordinary stellar black holes, provided that (i) the star cluster is subjected to a massive inflow of gas
which rises the central dispersion velocity above a given threshold,
and that (ii) the stellar black holes present in the core had no time to be ejected by single-binary encounters prior 
the massive inflow. 
The lapse time between inflow events at the redshifts explored is $\lsim 0.1 $ Gyr.
Black hole ejection in star clusters are seen to occur typically on times longer than $\sim 1$ Gyr \citep{downing11}, so that
at early cosmic epochs this effect can be neglected. Conservatively, the results can be considered as upper limits.

We find that the GIRM channel is competitive, when compared to the channels studied in D09, D10 and D12 and can occur in concomitance 
with the formation of seeds from  other channels, and in particular with that from runaway collisions of stars, in young dense clusters
for which there is a partial overlap in the redshift space. We halt the simulation at $z\sim 6$, as our model to not account for
the black hole growth due to accretion. 
Typically the black holes have seed masses in excess of $300\,\msun$ with a mean of $\sim 1000\,\msun$, opening the possibility that these black holes may later grow as supermassive at redshifts as early as $z\sim 7$, typical of the most distant QSOs.

These masses are in the range of the so-called intermediate mass black holes, which have not been yet observed in the Universe. If these black holes do not remain confined in galaxy centres but become wandering (like in the case of minor mergers of the host halo with a larger one), they could reach $z=0$ without becoming massive black holes, and they could become potentially observable with their original mass.

There is no model yet that can describe the thermodynamics of  a massive  inflow of gas onto a pre-existing star cluster. The gas funnelled  toward the centre of the cluster likely cools and fragment into stars, forming a more massive nuclear cluster nested into a primitive, lighter  one.  
The key requirement of the DMB11 model, not proved to be realistic yet, is a steepening of the gravitational potential such to lead to a major enhancement of the velocity dispersion inside the core of stellar mass black holes to trigger their relativistic collapse.
Further modelling is necessary in this direction (Galanti et al. in preparation 2014).

\section*{Acknowledgments}
The authors are pleased to thank Jillian Bellovary, Melvin Davies, and Coleman Miller for a critical reading of the manuscript and
their enlightening comments, and the Referee whose comments and queries helped improving the manuscript. 
MC and MV thank the Kavli Institute for Theoretical Physics at UC Santa Barbara for hospitality
during progress on this work, while the program ``A Universe of Black Holes'' was
taking place. 
MV acknowledges funding support from a Marie Curie Career Integration grant (PCIG10-GA-2011- 303609) within the framework FP7/PEOPLE/2012/CIG. This research was supported in part by the 
National Science Foundation under Grant No. NSF PHY11- 25915, through the Kavli Institute for Theoretical Physics and its program ``A Universe of Black Holes''. 

\bibliographystyle{mn2e}
\bibliography{Biblio}
\label{lastpage}
\end{document}